
\input phyzzx
%
\catcode`\@=11
\paperfootline={\hss\iffrontpage\else\ifp@genum\tenrm
 -- \folio\ --\hss\fi\fi}
\def\titlestyle#1{\par\begingroup \titleparagraphs
 \iftwelv@\fourteenpoint\fourteenbf\else\twelvepoint\twelvebf\fi
 \noindent #1\par\endgroup }
\def\GENITEM#1;#2{\par \hangafter=0 \hangindent=#1
 \Textindent{#2}\ignorespaces}
\def\address#1{\par\kern 5pt\titlestyle{\twelvepoint\sl #1}}
\def\abstract{\par\dimen@=\prevdepth \hrule height\z@ \prevdepth=\dimen@
 \vskip\frontpageskip\centerline{\fourteencp Abstract}\vskip\headskip }
\newif\ifYUKAWA  \YUKAWAtrue
\font\elevenmib   =cmmib10 scaled\magstephalf   \skewchar\elevenmib='177
\def\YUKAWAmark{\hbox{\elevenmib
 Yukawa\hskip0.05cm Institute\hskip0.05cm Kyoto \hfill}}
\def\titlepage{\FRONTPAGE\papers\ifPhysRev\PH@SR@V\fi
 \ifYUKAWA\null\vskip-1.70cm\YUKAWAmark\vskip0.6cm\fi
 \ifp@bblock\p@bblock \else\hrule height\z@ \rel@x \fi }

\def\schapter#1{\par \penalty-300 \vskip\chapterskip
 \spacecheck\chapterminspace
 \chapterreset \titlestyle{\ifcn@@\S\ \chapterlabel.~\fi #1}
 \nobreak\vskip\headskip \penalty 30000
 {\pr@tect\wlog{\string\chapter\space \chapterlabel}} }

\def\ssection#1{\par \ifnum\lastpenalty=30000\else
 \penalty-200\vskip\sectionskip \spacecheck\sectionminspace\fi
 \gl@bal\advance\sectionnumber by 1
 {\pr@tect
 \xdef\sectionlabel{\ifcn@@ \chapterlabel.\fi
 \the\sectionstyle{\the\sectionnumber}}%
 \wlog{\string\section\space \sectionlabel}}%
 \noindent {\S \caps\thinspace\sectionlabel.~~#1}\par
 \nobreak\vskip\headskip \penalty 30000 }


\papers

\def\lkakko{\vbox{\vskip0.065cm\hbox{(}\vskip-0.065cm}}
\def\rkakko{\vbox{\vskip0.065cm\hbox{)}\vskip-0.065cm}}
\def\YUKAWAHALL{\hbox to \hsize
 {\hfil \lkakko\twelvebf YUKAWA HALL\rkakko\hfil}}

\def\figmark#1{\par\vskip0.5cm{\hbox{\centerline{
 \vbox{\hrule height0.5pt%
 \hbox{\vrule width0.5pt\hskip4pt%
 \vbox{\vskip5pt\hbox{Fig.#1}\vskip4pt}%
 \hskip4pt\vrule width0.5pt}%
 \hrule height0.5pt}}}}\par\noindent\hskip0.1cm\hskip-0.1cm}



\def\addeqno{\ifnum\equanumber<0 \global\advance\equanumber by -1
 \else \global\advance\equanumber by 1\fi}


\mathchardef\Lag="724C
\def\sqr#1#2{{\vcenter{\hrule height.#2pt
 \hbox{\vrule width.#2pt height#1pt \kern#1pt\vrule width.#2pt}
 \hrule height.#2pt}}}


\def\cref#1{\rlap,\attach{#1)}}
\def\ref#1{\attach{#1)}}



\newdimen\ex@
\ex@.2326ex
\def\boxed#1{\setbox\z@\hbox{$\displaystyle{#1}$}\hbox{\lower.4\ex@
 \hbox{\lower3\ex@\hbox{\lower\dp\z@\hbox{\vbox{\hrule height.4\ex@
 \hbox{\vrule width.4\ex@\hskip3\ex@\vbox{\vskip3\ex@\box\z@\vskip3\ex@}%
 \hskip3\ex@\vrule width.4\ex@}\hrule height.4\ex@}}}}}}
\def\txtboxed#1{\setbox\z@\hbox{{#1}}\hbox{\lower.4\ex@
 \hbox{\lower3\ex@\hbox{\lower\dp\z@\hbox{\vbox{\hrule height.4\ex@
 \hbox{\vrule width.4\ex@\hskip3\ex@\vbox{\vskip3\ex@\box\z@\vskip3\ex@}%
 \hskip3\ex@\vrule width.4\ex@}\hrule height.4\ex@}}}}}}
\newdimen\exx@
\exx@.1ex
\def\thinboxed#1{\setbox\z@\hbox{$\displaystyle{#1}$}\hbox{\lower.4\exx@
 \hbox{\lower3\exx@\hbox{\lower\dp\z@\hbox{\vbox{\hrule height.4\exx@
 \hbox{\vrule width.4\exx@\hskip3\exx@%
 \vbox{\vskip3\ex@\box\z@\vskip3\exx@}%
 \hskip3\exx@\vrule width.4\exx@}\hrule height.4\exx@}}}}}}

\chardef\fontD="1A

\catcode`@=12

\def\nref#1{${}^{#1}$)}

\pubnum={YITP/K-997}
\date={December 1992}
\titlepage

\vskip 10mm
\centerline{\fourteenbf Poisson and Porter-Thomas Fluctuations}
\centerline{\fourteenbf in Off-yrast Rotational Transitions}
\vskip 5mm
\vskip 10mm
\vskip 15mm
\centerline{M.Matsuo${}^{a)}$, T.D\o ssing${}^{b)}$, B.Herskind${}^{b)}$
and S. Frauendorf${}^{\ c)}$}

\vskip 15mm
\centerline{\it a) Yukawa Institute for Theoretical Physics, Kyoto University,
Kyoto 606-01, Japan}
\centerline{\it b) Niels Bohr Institute, University of Copenhagen,
DK-2100 Copenhagen \O , Denmark}
\centerline{\it c) Institut f\"ur Kern- und Hadronenphysik
FZ-Rossendorf, PF 19, O-8054 Dresden, Germany}

\vskip 40mm
\line{\bf Abstract \hfill}
{ 
Fluctuations associated with
stretched E2 transitions from high spin levels
in nuclei around $^{168}$Yb are investigated by
a cranked shell model extended to include residual two-body interactions.
It is found that the gamma-ray
energies behave like random variables and the energy
spectra show the Poisson fluctuation,
in the cranked mean field model
without the residual interaction.
With two-body residual interaction included, discrete transition pattern
with unmixed rotational bands is still valid up to around
600 keV above yrast, in good agreement with experiments.
At higher excitation energy,
a gradual onset of rotational damping emerges. At 1.8 MeV above
yrast, complete damping is observed with  GOE type
fluctuations for both energy levels and transition strengths
(Porter-Thomas fluctuations).
}

\vfill
\break
\line{\bf 1. Introduction \hfil}

Recently, improved detectors and analysis techniques
have focused the attention on the
off-yrast nuclear structure of deformed nuclei at high spin.
For example, it is  expected that
the rotational band structure typical of the near-yrast regions
is smeared out
at an intrinsic excitation energy of order of $E_x \gsim 1$ MeV
measured from the yrast in rare-earth nuclei \nref{1}.
There are also arguments that
this damping property of the rotational motion is related to
onset of chaotic behavior in deformed nuclei \nref{2,3,4}.

Information of the nuclear structure of rotating nuclei
is usually obtained from identifying the E2 transitions
in rotational bands near yrast, while
the gamma-rays emitted from levels at higher excitation energy form
the so-called ``quasi-continuum" spectra in which
individual transitions are not distinguished because of the
high level density and finite detector resolution.
The fluctuation analysis
technique, which has been developed recently \nref{5}, is useful for
a study of the quasi-continuum spectra.
{}From the analysis of fluctuations in the ridge region of
2-D $E_\gamma $-$E_\gamma $ spectra, an effective number of the E2-decay paths
associated with decay along rotational bands
is extracted.
The fluctuation analysis technique is based on
the assumption that the
gamma-ray energies of the rotational E2 transitions behave like
random variables, as if they are like rain drops
falling at arbitrary positions \nref{5}.
This is the key assumption in making
a connection between the fluctuation of counts in the
quasi-continuum spectra and the number of rotational bands
passed in the decay.

It should be emphasized that
we actually do not know the underlying fluctuation behavior of the
E2 transitions.
Consider rotational bands in the near-yrast region which  do
not interact strongly with each other. One may expect
that if some
dynamical quantum numbers or symmetries
persist, they imply regular patterns or correlations in the rotational
transition energies.
On the other hand, the rotational perturbation at high
spin may be able to destroy such regularity by breaking the
dynamical symmetries.
Hence the validity of
assumption of  random E2 transition energies,
adopted by the fluctuation analysis, may depend on whether
any dynamical symmetries in the rotating nuclei survive.

The E2 transitions are expected to show
very  different fluctuation
behavior when the rotational damping sets in. In this case,
the E2 transitions from an initial level fragment into many
final levels because many rotational bands with
different intrinsic configurations are strongly mixed.
One may readily expect that the wave functions of
the strongly mixed levels are very complicated, and
that strengths of individual components of the fragmented E2 transitions
strongly fluctuate,  reflecting the complexity of the mixed wave
functions. A description of the strength fluctuation
in the strongly mixed levels may be governed by the random matrix model
(the Gaussian Orthogonal Ensemble),
in which the strength fluctuations show the Porter-Thomas
distribution \nref{6}. It is not clear, however,
whether the Porter-Thomas strength
fluctuation is realized in the E2 transition matrix elements
in the region in interest, with
intrinsic excitation energy of a few MeV.

In this paper we examine theoretically
the fluctuations associated with  the E2 transitions in
rotating nuclei
on the basis of a model for
the high spin off-yrast levels.
In particular we shall investigate whether
randomness of the transition energies is realized
for the non-interacting
rotational bands,
and whether
the strengths
in the damped E2 transitions exhibit
Porter-Thomas fluctuations.

\vskip 5mm
\line{\bf  2. Extended cranked shell model \hfil}

The model we use is an extension of the cranked shell model,
in which we take into account not only the cranked single-particle
Hamiltonian  but also
the two-body residual interactions which cause the rotational damping.
As the single-particle fields we adopt
the Nilsson potential and the cranking term $-\omega J_x$,
but neglect the static pairing gaps,
since the levels emitting the quasi-continuum
gamma-rays are in the high angular momentum region around
$30\hbar$ to $60\hbar$,
where the pairing gap is expected to be negligible.

It is straightforward to introduce the independent-particle
configurations using  \break
single-particle orbits in the
cranked Nilsson field. Many-particle and many-hole ($n$p-$n$h) excitations
should be included to form a suitable basis
for the off-yrast levels. The independent-particle configuration
is hereafter denoted as $\ket{\mu(\omega)}$.
As an ansatz, it is assumed that the
$n$p-$n$h configurations correspond to the rotational bands which
would exists in the off-yrast region if there were no
interactions between the rotational bands.
A residual two-body interaction $H_{res}$ is assumed between
the rotational bands.
Thus the Hamiltonian in this model is expressed:
$$
H=h_{Nilsson}-\omega J_x + H_{res}    \eqno{(1)}
$$
as a function of the rotational frequency $\omega$ (using
the convention $\hbar=1$).
Details of the residual interaction are discussed in sect.4.

The Hamiltonian $H$ is diagonalized within
a suitably truncated basis set of
$\{\ket{\mu(\omega)}\}$. The diagonalization is
separately made
among the basis states with the same parity $\pi$ and
signature $\alpha$ at fixed $\omega$.
The resultant levels are denoted by $i$, their energy by
$E_i(\omega)$ and
their wave functions by $\ket{i(\omega)}$.
The routhians of the levels are
defined by
$e'_i (\omega ) = E_i (\omega )-E_{ref}(\omega )$,
where the reference energy  $E_{ref}(\omega )$ is chosen
as the lowest energy level in the numerical calculations.

The rotational stretched E2 transitions are
calculated as follows. As far as we consider  well-deformed nuclei
with stable shape, it may be assumed as a good approximation
that intra-band transitions dominate if no interaction between
the bands exists and the quadrupole moments are
the same (constant $Q_0$) for all the bands.
The E2 transition strength and the associated $\gamma$-ray
energy can be calculated as
$$
S_{ij} = g Q_0^2
\left\langle j(\omega -2/{\cal{J}})\vert i(\omega )\right\rangle^2
\ \ \ ,                                                       \eqno{(2)}
$$
$$
E_\gamma = 2\omega  + e'_i(\omega )-e'_j(\omega -2/{\cal{J}})\equiv
E_{ij} \ \ ,  \eqno{(3)}
$$
for a transition between levels $i$ and $j$. Here
$\ket{j(\omega -2/{\cal{J}})}$ is a solution of the
extended cranked shell model at the
rotational frequency $\omega -2/{\cal{J}}$, $\cal{J}$ denoting the average
moment of inertia.  In this way, $i$ and $j$ represent states
at angular momentum $I$ and $I-2$, respectively.
The factor $g$ in Eq.(2) represents the geometrical factor which
arises from the rotational $D$-function. Only
the overlap matrix element
$\left\langle j(\omega -2/{\cal{J}})\vert i(\omega )\right\rangle^2$
plays a role in the
following analysis since we always normalize the E2 strengths.

The above framework is similar to the one employed
by S. \AA berg \nref{3,4}
in many respects.
References \nref{3,4}, however, introduce an additional procedure to transform
the energy of the $n$p-$n$h basis $\ket{\mu(\omega)}$ to the lab-frame energy
as a function of the angular momentum $I$, and the diagonalization is
made at a fixed $I$. We do not follow this procedure here
because it is not expected to influence the
statistical properties of the results,
although individual levels may differ in both descriptions. The
procedure in refs.\nref{3,4} may be useful for direct comparison with
experiments, especially when describing $E_\gamma-E_\gamma$ spectra.
The procedure described above is a more straightforward theoretical
framework and simpler in practical numerical calculations.

With the residual interactions incorporated, the extended cranked
shell model can describe  both
the rotational bands near the yrast line and
the levels far off yrast, where strong
bands mixing may occur. Nevertheless, we shall
divide our analysis into two parts for convenience of
discussion. We first neglect the residual interaction, as in
the conventional cranked shell model. The
intra-band E2 transitions in the near-yrast regions
are then described by this approximation since
it is expected that the rotational bands in this region do not mix with
each other. We shall discuss the statistical properties
for those intra-band rotational transitions in sect. 3.
In sect. 4, we discuss the results of the
extended cranked shell model where the residual interactions are
incorporated, concentrating on
the E2 transitions in the far-off-yrast region where
the rotational damping is found.
In sect. 5 a brief discussion  of the
transitional region between unmixed and mixed rotational bands
is given.

\vskip 5mm
\line{\bf 3. Non-interacting rotational bands \hfil}

In the cranked shell model without  residual interaction,
the rotational bands are simply described as
independent-particle configurations
assigned to the cranked single-particle
orbits (called $n$p-$n$h bands hereafter)
and only the intra-band stretched E2 transition is considered.
The transition strength is almost constant,
irrespective of the configurations
of the bands, provided that the deformation is stable.
On the other hand, the transition energy
varies from band to band because of differences in the alignment.
Thus it is appropriate to focus our
attention on the fluctuation property of
the {\it transition energies}.

\vskip 5mm
\line{\tenrm 3.1. DISTRIBUTION OF $\gamma$-RAY ENERGIES \hfil}

The gamma-ray energy of the
the stretched intra-band E2 transitions of a $n$p-$n$h band is expressed as
$$ \eqalign{
    E_\gamma  & = 2 \omega _\mu (I)  \cr
              & \simeq
2\omega +{2\over\cal{J}}(I-J_\mu (\omega)) \cr
}            \eqno{(4)}
$$
in terms of the expectation value $J_\mu (\omega )$
of $J_x$ for an independent-particle
configuration $\mu $ at $\omega $. In expressing the rotational frequency
$\omega _\mu (I)$ which corresponds to spin value $I$,
we make the approximation as $I=J_\mu (\omega _\mu (I))
\simeq J_\mu (\omega )+{\cal{J}}(\omega _\mu (I)-\omega )$
with $\cal{J}$ being the moment of inertia.
Eq.(4) can also be derived from Eq.(3).
It should be noticed that the stretched E2 transitions
occur successively following the spin sequence
$\cdots I+2 \rightarrow I \rightarrow I-2 \cdots$.
In order to investigate the properties of spectra containing
all these sequential transitions we
displace the $E_\gamma $ spectrum by $\pm 4/\cal{J}$, $\pm 8/\cal{J}$, etc,
and fold all the displaced spectra. Here the energy unit $4/\cal{J}$ of the
displacement is assumed to be
0.06 MeV which is appropriate for well-deformed nuclei with $A \sim 170$.
The folded $E_\gamma $ spectra are
equivalent to the spectra  of $J_\mu (\omega )$ modulo 2, which we
denote by $\Delta J_\mu$.
A typical channel width of the experimental spectra
is about one tenth of the $\gamma $-ray energy unit $4/\cal{J}$.
Correspondingly, we calculate the distribution of
$\Delta J_\mu  = J_\mu (\omega ) \  {\rm modulo} \ 2$
after sorting the $\Delta J_\mu $'s into
10 channels. Figure 1 shows a typical result for ${}^{168}$Yb.
The Nilsson parameters in ref.\nref{7}
and the deformation parameters $\epsilon _2=0.255, \epsilon _4=0.014$
in ref.\nref{8}
are used.
Here the spectra in fig.1 are
calculated for the lowest 20 bands (fig.1(a)) with the signature and
parity quantum numbers $(+,0)$ as well as  for the next 20 bands,
the third 20 bands
and so on, altogether 10 bins of bands.

\figmark{1}

\vskip 5mm
\line{\tenrm 3.2. FLUCTUATIONS OF DISCRETE $\gamma$-RAY SPECTRA \hfill}

The spectra in fig.1 display a quite random fluctuation.
In order to analyze the fluctuation quantitatively,
the second moment or the variance of the
spectra is calculated
in accord with the fluctuation analysis of the
experimental data \nref{5}.
The second moment $\mu _2$ here is defined
by $\mu _2=(\sum_n m_n^2)/N_{channel} - ((\sum_n m_n)/N_{channel})^2$ where
$m_n$ is the number of $\Delta J_\mu $'s in the $n$-th channel.
{}From the $\mu_2$'s obtained for the 10 bins we calculate the
average value at different  rotational
frequencies $\omega$. The result is plotted in fig.2
as a function of $\omega $.  For comparison we show
in the same figure the expected limit if
the distribution of  $\Delta J_\mu $ is completely random, i.e.
obeys a Poisson distribution.
The random limit is calculated by assuming that the
distribution of number $m_n$ in channel $n$ is binomial.
The resulting average value of $\mu _2$ is $(1-N_{channel}^{-1})\mu_1=1.8$
with the mean number $\mu_1=2$, and the statistical sampling allows a
deviation within 0.26. The standard deviation of $\mu_2$ in the limit is
$\sigma(\mu_2)=\sqrt{2 / N_{channel}}\ \mu_1=0.89$.
The random limit is indicated in
fig.2 with the horizontal line and the expected deviation with the
dashed lines.

\figmark{2}

It is clearly seen in fig.2
that the fluctuation in the calculated spectra follows the
random limit for rotational frequencies
$\omega $  higher than 0.3 MeV. It is also seen that
there is a systematic deviation from the limit for
low rotational frequencies. We find that this deviation arises from
the fact that time-reversal doublets in the cranked single-particle
orbits are less dissolved for lower $\omega $, because the
rotational perturbation is weak.
This causes a degeneracy in $\Delta J_\mu $, which results in larger
fluctuations.
It should be noted, however, that the cranked shell model does not take
into account effects which influence the spectra at low spin.
For example, the intrinsic angular momentum components
along the symmetry axis ($K$ components)
is expected to split the degenerate
states which have different $K$ values.
The effect of the $K$ component on the $\gamma$-ray energy is evaluated
according to Eq.(A.5) given in the appendix. We estimate  values of $K$ for
rotational bands by assigning them to those that are
obtained at $\omega=0$.
This gives a order-of-magnitude estimate of the $K$ effect.
The resulting second moments of
the $\gamma$-ray energy spectra are shown in fig.2 as triangles.
One can see that the inclusion of the K effect enhances
the tendency toward the random limit.
We conclude that the random limit
is achieved for the high spin region
$\omega  \gsim 0.3$ MeV or
$E_\gamma  \gsim 0.6$ MeV.

It is  neccesary to check the above conclusion because
our sampling of the rotational bands in the above analysis may be
somewhat artificial. We sample the lowest 200 $n$p-$n$h bands for each
$(\pi,\alpha)$ while
the number of the unmixed rotational bands
is order of a few tens
according to the experimental fluctuation
analysis \nref{5}.
We therefore perform a  physically more relevant analysis by sampling
only the bands near yrast.
{}From $n$p-$n$h bands with all combinations of $(\pi,\alpha)$, the
lowest 20 bands are picked up.
Associated
$\Delta J_\mu$'s
are sorted into the channels
in the same manner as in fig.1. To obtain a set of the
second moments $\mu_2$  of spectra,
we perform the same calculation for 10 even-even
nuclei around ${}^{168}$Yb, namely,
${}^{162-166}$Er,${}^{164-170}$Yb, and ${}^{168-172}$Hf.
{}From the set of $\mu_2$'s, we
calculate the mean value and the
standard deviation $\sigma (\mu_2)$ as a function of $\omega$. The result is
shown in fig.3, also with the $K$ effect  taken into account.
{}From the comparison with the average $\mu_2$ and $\sigma(\mu_2)$
of the random Poisson distribution,
the following picture emerges: {\it The
intra-band $\gamma$-ray energies of the rotational bands in the
independent-particle cranked shell model appear to be randomly
distributed,  at least at high spin}.

\figmark{3}

\vskip 5mm
\line{\bf 4. Mixed rotational bands \hfil}

\line{\tenrm 4.1. RESIDUAL INTERACTION \hfil}

The two-body residual interaction consists of
the monopole and quadrupole
pairing interaction and the quadrupole-quadrupole interaction,
which represent large components of the two-body
matrix elements. We also introduce additional components of
the two-body matrix elements, which may arise
from the other multipolarities. By assuming the additional
components are structureless, we give them simply in terms of
random numbers following the Gaussian distribution
around zero mean value.
The strength of the residual interaction is
fixed as follows. For the monopole
pairing interaction we adopt a standard value
$ G_0 = 20/A$ MeV. We fix the quadrupole pairing strength
by $G_2/G_0 = 35/3R^4$ which is derived from the multipole
expansion of the delta-force\nref{9}.
The quadrupole-quadrupole interaction strength
is given by the selfconsistent values \nref{10}
for both iso-scalar and iso-vector
components with the polarization effect
($\chi _{nn}=\chi _{pp}=0.7\chi _{self}$,
$\chi _{np}=3.3\chi _{self}$ where the $\chi _{self}$
is the selfconsistent value).
In the numerical calculation, we
used the stretched quadrupole operator (instead of the ordinary ones)
for  practical reasons.

The strength of the random two-body
matrix elements is fixed as in ref.\nref{1}, namely the
spreading width
$\Gamma^{\downarrow}_\mu = 0.039 \left({A\over 160}\right)^{-1/2}E^{3/2}$
MeV of independent-particle configuration
is extracted from the systematics of the single-particle spreading width,
and related to the random two-body interaction through the
Fermi golden rule $\Gamma^{\downarrow}_\mu =2\pi \rho _{{\rm 2-body}}v^2$.
The density $\rho _{{\rm 2-body}}$ of levels interacting through the two-body
interactions is calculated explicitly in the cranking model without
the interactions and it is  parametrized as
$ \rho_{{\rm 2-body}} \simeq 22E^{3/2}$ MeV${}^{-1}$.
The root-mean-square value $v$ of the random
two-body matrix elements is thus fixed to
$v=16$ keV.

It can be argued that there is an uncertainty in the residual interaction
strength. However this uncertainty is not crucial in the
following analysis as long as we do not aim at very accurate numbers in the
excitation energy. As is discussed in
ref. \nref{3}, the two-body interaction strength scales proportionally to
the spacing between interacting configurations, i.e.
$d_{{\rm 2-body}} = 1/\rho_{{\rm 2-body}} \propto E^{-3/2}$. If the
scaling is a good approximation,
a factor 2, for example,  in the interaction strength
results in a scaling of $2^{2/3} \sim 1.6$  in the excitation energy.
This should, however, only be taken as a guideline since we are
investigating properties concerning the full set of states.

\vskip 5mm
\line{\tenrm 4.2. CHOICE OF BASIS \hfil}

In diagonalizing the
Hamiltonian matrix
the basis of $n$p-$n$h configurations is truncated
in the following way. First,
we construct all the $n$p-$n$h configurations that lie below
$E_x = 3.0$ MeV above the lowest energy configuration. Next the
diagonal energies due to the residual interaction are
calculated for these
configurations.
The basis of the diagonalization
consists of the lowest configurations including the diagonal interaction
up to a certain number, eg, 600, 1000, or 1400, for each $(\pi,\alpha)$.
The levels at $\omega-2/\cal{J}$ are calculated using the same basis
as defined at $\omega$.
Figure 4 shows the density of levels calculated thus for the
levels with $(-,1)$ in ${}^{168}$Yb at $\omega=0.5$ MeV.
Results with the three different truncations are plotted to
display the convergence. The basis of 1000 states is found to be
sufficient for describing
the levels up to 2.0 MeV from the yrast, amounting to over 200 levels
for each $(\pi,\alpha)$.
In the following the calculation with 1000 states is discussed.
Most of the figures presented below are for calculation on
${}^{168}$Yb at
$\omega=0.5$ MeV, although a few include the results from
the 10 neighboring even-even nuclei
${}^{162-166}$Er,${}^{164-170}$Yb, and ${}^{168-172}$Hf.

\figmark{4}

\vskip 5mm
\line{\tenrm 4.3. TRANSITION STRENGTH FLUCTUATIONS \hfil}

When the residual interaction and band mixing are included
in the calculations, we see that
rotational transitions from each level branch  out into many final
levels.
A clear evidence of this manifestation of rotational damping
is shown in fig.5,
which depicts four
typical examples of the calculated rotational
strength distributions plotted
as a function of $\gamma$-ray energy. The strengths shown
in each quarter correspond to decay  from  individual levels.
The four levels with $(\pi,\alpha)=(-,1)$
from different regions of excitation energy
demonstrate  how the rotational decay will change its structure as the
excitation energy is increased.

\figmark{5}

For the yrast level, the rotational strength concentrates into
almost only one transition, indicating the dominance of the
rotational band structure. For the 6-th level
at $E_x=758$ keV, the strength fragments into several
pieces. We may consider  this fragmentation as
damping of the rotational transitions, although this is  rather subtle
because of the small number of fragments.
The situation becomes clearer for the 24-th level ($E_x=1.133$ MeV)
and for the 200-th level ($E_x=1.948$ MeV): the rotational strength
from these levels splits into a large number
of fragments and there appears
a smooth profile from which a damping width can be extracted.
Thus we see in fig. 5 the onset of rotational damping
around several hundred keV.

Next, let us focus our attention on the
statistical properties of the rotational
transitions, in particular, on those associated with the rotational
damping. As seen in fig.5, the height of the peaks
or magnitude of the rotational strengths fluctuates significantly
around the smooth profile of the strength distributions.
This fluctuation of the strength is what would be expected
for levels with
complex mixing. To clarify the nature of the strength fluctuation,
we shall discuss it on a quantitative basis. In particular, we shall
discuss whether it exhibits the Porter-Thomas fluctuations or not.

In analyzing the fluctuation in the rotational strength it should be
noted that the strength distribution exhibits an overall smooth profile,
which may be characterized by the central peak position
and the rotational damping width (or FWHM).
The strength fluctuation
seen in fig.5, therefore, should be measured relative
to the smooth profile.
It is appropriate to single out the relative fluctuation
by normalizing the E2 strengths with respect to
the smooth profile. For a transition from $i$ to $j$,
the normalized strength is defined by
$$
s_{ij} = {S_{ij} \over \left\langle S_{ij}\right\rangle} \ \ ,  \eqno{(5)}
$$
where $\left\langle S_{ij}\right\rangle$
represents the strength in the absence of
fluctuations. By definition, the normalized strength
should have unit mean. The smoothed strength $\left\langle S_{ij}\right\rangle$
should be a function of the $\gamma$-ray energy $E_{ij}$ and
the excitation energy of the levels. One can evaluate it by
$$
\left\langle S_{ij}\right\rangle
= \overline{S}(E_{ij})/\rho (e'_j) \ \ ,     \eqno{(6)}
$$
where $\overline{S}(E)$ is the smooth profile of the
rotational strength distribution function.
The smooth strength function $\overline{S}(E)$ is extracted by taking a
Strutinsky-type smoothing of the microscopically calculated strength
distribution
$$
            S(E) = \sum_{ij}S_{ij}\delta (E-E_{ij})/\sum_{i}  \ \ ,
    \eqno{(7)}
$$
where the average over initial levels yields a reliable
smoothing. In the Strutinsky smoothing, a Gaussian width of 50 keV
is used. The level density $\rho(e)$ is determined by a
fitting to the constant temperature formula\nref{12}
$\rho(e)=T^{-1} e^{(e-E_0)/T}$.

\figmark{6}

In order to see the excitation energy dependence, the initial levels
with the same $(\pi,\alpha)$ are grouped into bins,
each of which contains 50 levels.
The excitation energies  of these bins are approximately
$0.0\ -\ 1.4$ MeV for $1\ -\ 50$-th levels,
$1.4\ -\ 1.7$ MeV for $51\ -\ 100$-th,
1.8 MeV for $101\ -\ 150$-th, and 1.9 MeV for $151\ -\ 200$-th. The rotational
strength distribution calculated for these bins are shown in
fig.6 as well as the smoothed strength function $\overline{S}(E)$
plotted with the solid curves.
We sample the normalized strengths of the transitions
whose $\gamma$-ray energy $E_{ij}$
lies within the interval of the full width at the one eighth of maximum of
$\overline{S}(E)$. For the lowest bin, for example, those transitions
picked up satisfy 881 keV $< E_{ij} < 1170 $ keV.

\figmark{7}

After the
sampling we calculate  distribution of the
normalized strength $s_{ij}$. The result is shown in
fig. 7 , together with a  Porter-Thomas distribution
$$
P(s)ds = (2\pi s)^{-1/2}e^{-s/2}ds                 \eqno{(8)}
$$
for comparison. It is clear from fig.7 that
that the distribution of the
normalized strengths approaches  that of Porter-Thomas for increasing
excitation energy and almost reaches it in the
region $E_x \gsim 1.8$ MeV.  Deviation from the Porter-Thomas
is seen in the lowest bin which corresponds to $E_x=0.0 - 1.4$ MeV
(also slightly in the next bin).

\figmark{8}

In the above analysis it has implicitly been assumed that the
property of the strength fluctuation does not depend on the
$\gamma$-ray energy. Let us show this for the highest bin
which clearly displays
Porter-Thomas fluctuations. For this purpose we subdivide
the samples of the normalized strength into two groups according to
$\gamma$-ray energy. One subgroup is for the transitions whose
$\gamma$-ray energies are close to the central peak of the
rotational strength function. In practice we choose
the transition energies within the Full-Width-Half-Maximum of the
smoothed strength function. The other
is for transition energies which lie in the tail portion of the
strength function (between the half and one eighth of the maximum value).
Figure 8 shows the result of this analysis
for transitions in the
fourth bin ($151 - 200$-th levels at $E_x \sim 1.9$ MeV in ${}^{168}$Yb)
of figs.6 and 7. Regardless of whether the transitions are
from the central or from the tail portions,
the strength fluctuation is found to obey the Porter-Thomas distribution.

It should be noticed here that the Porter-Thomas distribution
manifest itself only for the relative strengths which are
normalized with respect to the smooth profile of the
rotational strength function. This explains why the Porter-Thomas
distribution emerges here while it does not in the analysis made
in ref.\nref{4}.

\vskip 5mm
\line{\tenrm 4.4. BRANCHING NUMBER \hfil}

Let us analyze the strength fluctuation again, but this time
by looking into it in terms of a quantity which is closely related
to the experimental observables.
In the experimental fluctuation analysis \nref{5}
the number of the rotational decay paths $n_{path}^{(2)}$
is extracted from the second moment
of the spectral fluctuation.
Assuming one-dimensional spectra (or observing one-step transitions),
the number of paths is given by
$$
{1\over n_{path}^{(2)}} =
\sum_i {f_i^2 \over n_{{\rm branch}}(i)} \ \ ,  \eqno{(9)}
$$
$$
n_{{\rm branch}}(i) =  \left( \sum_j w_{ij}^2 \right)^{-1} \ \ ,   \eqno{(10)}
$$
where $f_i$ represents the probability for flow passing through
the level $i$, and $w_{ij}$ the probability for the
rotational transition to pass from
level $i$ to level $j$ (with normalization
$\sum_j w_{ij} =1$). The quantity $n_{{\rm branch}}(i)$ is a measure
of the
number of transitions which branch out of level
$i$ (or simply the branching
number). With the microscopic model
the branching numbers are determined by the transition probability
as given by
$$
w_{ij} = S_{ij} / \sum_j S_{ij} \ \ ,                      \eqno{(11)}
$$
where the energy dependent factor $\propto E_\gamma^5$ is neglected.
On the other hand, the average branching number expected for
a group of levels whose E2 strength displays
Porter-Thomas fluctuations can be estimated analytically by
using a suitable profile
of the rotational strength function. In fact, the Gaussian form
with FWHM of the rotational damping width $\Gamma_{rot}$ is adopted. The
average branching number in the Porter-Thomas limit is then given by
$$ \eqalign{
\overline{n_{{\rm branch}}(i)_{\rm,PT}}
& = {1\over6}\sqrt{2\pi\over \ln 2}
\ \Gamma_{rot}\ \rho(E_i) \cr
  & \approx 0.5 \ \Gamma_{rot} \ \rho(E_i) \ \ . \cr
} \eqno{(12)}
$$

\figmark{9}

Figure 9 compares the Porter-Thomas limit Eq.(12)
with the average branching number
calculated microscopically by means of Eq.(10).
In Eq.(12) we adopt $\Gamma_{rot}= 200$ keV.  This value
agrees well with the FWHM's  extracted from
the smoothed strength function shown in fig.6. Actually
they are approximately 170 keV and the agreement is fulfilled
within an accuracy of
15 $\%$. The validity of Eq. (12) supports  again that the
Porter-Thomas fluctuation is realized in the damped rotational
transitions.

\vskip 5mm
\line{\tenrm 4.5. ENERGY LEVEL FLUCTUATIONS \hfil}

It may be interesting to look into the fluctuation in the
energy levels because it is expected to show the properties
of the GOE random matrix theory if the Porter-Thomas strength
fluctuation is a consequence of the complex mixing of the
rotational bands. Here we test the $\Delta_3$ statistics
\nref{6} as a measure of the energy level fluctuations.
The calculated spectra are unfolded for each $(\pi,\alpha)$
by use of a fitted
level density, parametrized according to
the constant temperature formula.
An example of the fitting is shown in the inset of fig. 10.
The mean value of $\Delta_3$ is calculated
as a function of the interval length $L$ (the
unit is the average spacing).
As in the preceding section, we subdivided the levels into
bins of 50 levels.

\figmark{10}

Figure  10 shows that the $\Delta_3$ statistics approaches
the GOE limit with increasing excitation energy  and that
the limit is reached for the levels above $\#150$ at
$E_x \sim 1.8$ MeV.
It is remarkable, compared to the results shown in fig.7,
that the tendency toward the GOE limit is
completely consistent with that of the strength fluctuation.
Thus it can be concluded, from both the strength fluctuations and
the energy level fluctuations, that the mixing of the rotational bands
is as complex and random in the energy region $E_x \gsim 1.8$ MeV
as described by the GOE random matrix theory.
Concerning the energy level fluctuations, the
results are  consistent with the
work  by S. \AA berg \nref{3,4}.

We would like to emphasize the important implication of the above result
for the rotational transitions in the damping region.
A  spectrum which shows GOE fluctuations is known to be
``rigid" with very small fluctuations.
As a consequence, the fluctuations in transition energy
play a relatively unimportant role in the damping region, while
the strength fluctuation becomes the dominant origin of the
spectral fluctuations \nref{13}.

\vskip 5mm
\line{\bf 5. Onset of damping \hfil}

As discussed in connection with fig.5,
the model predicts both the damped transitions and
the discrete rotational transitions which characterize unmixed
rotational bands near yrast.
Thus it allows us to investigate the onset of the rotational damping
and also to analyze fluctuation properties of the discrete
rotational transitions, which was discussed in the preceding
section without including the residual interaction.

\figmark{11}

\vskip 5mm
\line{\tenrm 5.1. UNMIXED ROTATIONAL BANDS \hfil}

Figure 11 illustrates the behavior of rotational
transitions in the energy region where the damping is about to
set in. The
branching numbers calculated for  individual levels
are also shown in the figure.
A quantitative criterion for defining levels associated with unmixed
rotational bands may be given by a condition that
the branching number of a level is less than 2.
Accordingly the onset of the damping will be marked by
$n_{\rm branch} > 2$. If this criterion is assumed we
count 22 unmixed rotational bands  in $^{168}$Yb, and 30.9 bands
on average for the 10 rare-earth nuclei around spin
$42\hbar$. Considering  that
the pure rotational bands carry  essentially 100 $\%$ strength
of the intra-band transition, one may also define unmixed rotational-band
levels as those whose strongest transition has a large fraction
of the total strength, e.g. $1/\sqrt{2}=71\%$. This gives us
gives us 19 rotational bands  for $^{168}$Yb and 28.3
if averaged over the 10 neighboring nuclei.
The two criteria give almost the
same number. The above number of the unmixed rotational-band
levels is consistent with the experimental fluctuation analysis \nref{5},
which reports approximately 30 bands in $^{168}$Yb.

Since the residual interaction effect was not included
in the preceding analysis using the independent-particle
cranked shell model in sect. 3, we re-examine the fluctuation property
of the transition energies associated with the
discrete rotational transitions defined by the above criterion.
For this purpose, we select the lowest
20 levels which satisfy $n_{\rm branch}<2$ and use their
strongest transitions. The
second moment $\mu_2$ which represents
the fluctuation of the selected transition energies is calculated
for each nucleus at $\omega=0.5$ MeV
in the same way as in sect. 3. The average value
of $\mu_2$'s
over the 10 nuclei is found to be 2.16,
and their standard deviation is 0.80. These numbers
compare well with the random Poisson limits, 1.80 and 0.89, respectively.
Thus the residual interaction does not modify the conclusion
of sect.3, concerning the randomness of transition energies.

\vskip 5mm
\line{\tenrm 5.2. TRANSITION STRENGTH FLUCTUATIONS \hfil}

The energy for the onset of the rotational damping may be
defined although the damping sets in rather
gradually with increasing excitation energy  as seen in fig. 11.
{}From fig. 9, which shows average excitation energy
dependence of the branching number, and with the criterion
$n_{\rm branch}>2$,
the onset energy for $^{168}$Yb can be estimated to be
approximately $E_x \approx 600$ keV
above yrast.

Above the onset energy, the transitions from each level begin
to branch
into many small pieces and strong fluctuations in the strengths of the
fragmented transitions set in. Although in fig.7 almost pure
Porter-Thomas
fluctuations are found  in the energy region $E_x \gsim 1.8$ MeV,
a clear deviation from Porter-Thomas is found
for lower excitation energies. To describe the  energy
dependence of the strength fluctuation more quantitatively, the
lowest 50 levels are subdivided into smaller bins
of 10 levels and the distribution of the normalized strengths
is calculated for each bin, producing the result shown in fig.12.
In this figure, results for 10 nuclei are included to obtain good enough
statistics. It is seen in fig. 12 that deviations
from Porter-Thomas fluctuations are
more significant at lower excitation energies. Strong deviations are
found not only in
the lowest bin ($E_x \simeq 0.0 - 0.9$ MeV) which
contains the unmixed rotational-band levels but also in the
other bins where the rotational damping becomes dominating.
This indicates the presence of a transient region between  0.6 MeV
and 1.8 MeV where the
rotational transitions are already fragmented significantly, while
the strength fluctuations have not yet reached the Porter-Thomas limit.

\figmark{12}

\vskip 5mm
\line{\tenrm 5.3. LEVEL SPACING STATISTICS FOR UNMIXED BANDS \hfill}

{}From the viewpoint of discrete spectroscopy \nref{14}, fluctuation
properties in the energy levels near the yrast line are of interest,
although the spin region in the present analysis is higher
than current discrete spectroscopies can reach.
Figure 13 displays the nearest neighbor level spacing
distribution for the rotational-band levels near yrast.
As a reference, a result for
cranked shell model without the residual interaction is also shown.
In this figure only the lowest
energy levels satisfying $n_{\rm branch}<2$
(the lowest 5 levels for the independent-particle model) are sampled
for each $(\pi,\alpha)$ from the 10 nuclei. This gives 121 spacings
in the case of two-body interaction included, and
160 spacings for the independent-particle model.
The energy dependence of the fitted level density is
taken into account in normalizing the spacings (cf. sect.4.5 and fig.10).
The result
for the independent particle model resembles
the Poisson distribution.
With the residual interaction
included,
on the other hand,
there are less small spacings
than expected from the Poisson distribution.
The level spacing distribution appears to lie
in between the Poisson and the GOE limits.

\figmark{13}

\vskip 5mm
\line{\bf 6. Conclusion \hfil}

The E2 transitions in the
off-yrast region of well-deformed rare-earth nuclei are investigated
on the basis of an extended cranked shell model which
describes the $n$p-$n$h bands as well as the mixing among them.
Using the cranked shell model without
the residual interaction, the fluctuation in the gamma-ray energies
of the rotational transitions
associated with the
near-yrast non-interacting rotational bands are discussed.
It is  found
that the gamma-ray energy spectra at high spin show Poisson fluctuations
as if the gamma-ray energies were random variables.
The model with the residual interaction is found to
give rise to the damping of the rotational transition
above an excitation energy ($E_x\simeq 600$ keV).
It is seen that the magnitude of
the E2 strength fluctuates for
the fragmented components of the damped
rotational transitions.
Further up in excitation energy ($E_x \gsim 1.8$ MeV),
the
strength fluctuations are found to obey the Porter-Thomas distribution.
In the same energy region,
it is also found that the energy level fluctuations
distribute according to the GOE random matrix model
implying that the mixing of rotational bands is as complex
as in the random matrix theory. Using the properties
of the random matrix theory as a criterion for the
chaotic behavior of quantum systems \nref{15},
this region
appears to be chaotic. The energy region just above
the onset of damping appears as a transient region
where the strength fluctuation show significant deviations from the
Porter-Thomas distribution.

In the model we adopted, the boundary energy for the
onset of rotational damping is about 600 keV,
and the boundary for sharing full Porter-Thomas strength fluctuation is
around 1.8 MeV. These numbers, however, should be considered with
reservation since  they may depend on the residual interaction
as well as its strength. It is therefore important to extract
these energies in more detailed experiments.

On the theoretical side, it seems
interesting to study how the situation changes when
more realistic residual interactions are used.
Another interesting aspect of
band mixing concerns the correlations in the wave functions,
which extend
over several steps in angular momentum. These correlations are relevant
for a more precise comparison to two and higher fold experimental
spectra.
Such investigations are in progress.

\vfill
\break

\baselineskip 6mm

\line{\bf References \hfill}
\item{1)} B.Lauritzen, T.D\o ssing and R.A.Broglia, Nucl. Phys.
{\bf A457}(1986) 61.
\item{2)} T.Guhr and H.A. Weidenm\"uller, Ann. Phys. {\bf 193}
(1989) 489.
\item{3)} S.\AA berg, Phys. Rev. Lett. {\bf 64}(1990) 3119.
\item{4)} S.\AA berg, Prog. Part. Nucl. Phys. vol.28 (Pergamon 1992) p.11.
\item{5)} B.Herskind, A.Bracco, R.A.Broglia, T.D\o ssing,
A.Ikeda, S.Leoni, J.Lisle, \break
M.Matsuo, and E.Vigezzi, Phys. Rev. Lett.
{\bf 68}(1992) 3008. \hfil \break
B.Herskind, T.D\o ssing, S.Leoni, M.Matsuo, and E.Vigezzi,
Prog. Part. Nucl. Phys. Vol.28 (Pergamon 1992) p.235.
\item{6)} C.E. Porter and R.G. Thomas, Phys. Rev. {\bf 104}(1956)483.
\hfil\break
T.A.Brody, J.Flores, J.B.French, P.A.Mello,
A.Pandy and  S.S.M.Wong, Rev. \break Mod. Phys. {\bf 53}(1981) 385.
\item{7)} T.Bengtsson and I.Ragnarsson, Nucl. Phys.
{\bf A436}(1985) 14.
\item{8)} R.Bengtsson, S.Frauendorf and F.-R.May,
Atomic Data and Nuclear Data Tables, {\bf 35}(1986) 15.
\item{9)} I.Hamamoto, Nucl. Phys. {\bf A232}(1974) 445.
\item{10)} A.Bohr and B.R.Mottelson, Nuclear structure vol.2
(Benjamin 1975).
\item{11)} S.\AA berg, Nucl. Phys. {\bf 477} (1988) 18.
\item{12)} A.Gilbert and A.G.W.Cameron, Can. J. Phys. {\bf 43} (1965) 1446.
\item{13)} P.G.Hansen, B.Jonson and A.Richter,
Nucl. Phys. {\bf A518} (1990) 13.
\item{14)} J.D. Garrett, J.R. German, L. Courtney, and J.M. Espino,
Proc. Symp. on Future Directions in Nuclear Physics with 4$\pi$ Gamma
Detection Systems of the New Generation, Strasbourg, 1991, ed.
J. Dudek and B. Haas (American Institute of Physics, 1992) p.345.
\item{15)}
O. Bohigas, M.J. Giannoni and C. Schmit, Phys. Rev. Lett.
{\bf 52}(1984) 1. \hfil\break
O. Bohigas and H.A. Weidenm\"uller,
Ann. Rev. Nucl. Part. Sci. {\bf 38} (1988) 421.

\vskip 10mm
\line{\bf Acknowledgments \hfil}
One of the authors, MM, acknowledges the Nishina
Memorial Foundation for supporting his stay at
the Niels Bohr Institute, where parts of
this work has been carried out.
Illuminating discussions with S. \AA berg are gratefully
acknowledged.

\vfill
\break

\baselineskip 7mm


\line{\bf Appendix \hfil}

With the presence of a  $K$ component of the angular momentum,
the rotational frequency along the rotational axis can be defined by
$$
\omega(I)  =  {E(I+1)-E(I-1)\over I_x(I+1)-I_x(I-1)} \ \ , \eqno{(A.1)}
$$
$$
I_x(I)  =  \sqrt{I^2 - K^2} \ \ .  \eqno{(A.2)}
$$
For a $n$p-$n$h band in the extended cranked shell model, in which
the angular momentum $I_x$ along the rotational axis is calculated
as a function of $\omega$ as $I_x = J_\mu(\omega)$, Eq.(A.2) is used to
determine $\omega$ which corresponds to the angular momentum $I$.
$$
\sqrt{I^2-K_\mu^2}  = J_\mu(\omega_\mu(I))
  \simeq  J_\mu(\omega) + \cal{J} (\omega_\mu(I)-\omega) \ \ , \eqno{(A.3)}
$$
where $K_\mu$ is the $K$ value for the band $\mu$ and $\omega$ is a reference
frequency. Eq.(A.1) reduces to an equation which gives the associated
$\gamma$-ray energy,
$$
E_\gamma  \simeq  2\omega_\mu(I){\partial I_x \over \partial I}
   =  2\omega_\mu(I){I \over \sqrt{I^2-K^2_\mu}} \ \ . \eqno{(A.4)} $$
Combining (A.3) and (A.4), we get
$$ \eqalign{
E_\gamma & =  2 \omega {I \over \sqrt{I^2-K^2_\mu}} +
 {2 \over \cal{J}}\left(I-J_\mu(\omega){I \over \sqrt{I^2-K^2_\mu}}\right) \cr
& \simeq  2 \omega {\sqrt{J_\mu(\omega)^2+K_\mu^2}\over J_\mu(\omega)} +
{2\over \cal{J}}\left(I-\sqrt{J_\mu(\omega)^2+K_\mu^2}\right) \ \ .   \cr
} \eqno{(A.5)} $$
In deriving the final expression, we neglect the $I$-dependence in
the geometrical factor $I/\sqrt{I^2-K_\mu^2}$ by replacing it
with $\sqrt{J_\mu(\omega)^2+K_\mu^2}/ J_\mu(\omega)$. Eq.(A.5) can
be used in place of Eq.(4).

\vfill
\break
\line{\bf Figure captions \hfill}

\item{Fig.1}
The spectra of $\Delta J_\mu  = J_\mu $ modulo 2
of the basis $n$p-$n$h configurations
for $(\pi ,\alpha )=(+,0)$ bands in ${}^{168}$Yb
at $\omega =0.5$ MeV.
The first bin contains the first 20 bands in order of
the routhian energies,
the second bin is for the next 20 bands, and so on.

\item{Fig.2}
The averaged second moment $\mu _2$ of fluctuation
defined for the $\Delta J_\mu $ spectra of the $n$p-$n$h
basis bands with
(+,0) in ${}^{168}$Yb
as a function of the rotational frequency $\omega $, plotted
with closed circles.
The value at $\omega =0.5$ MeV
corresponds to the spectra shown in  Fig.1.
The triangles are calculated for the spectra in which
the effect of the $K$ components on the $\gamma$-ray energy is
taken into account.
The random (Poisson) limit is indicated by the horizontal line.
The dashed lines represents the uncertainty expected for
the average $\mu_2$ in the random limit.

\item{Fig.3} The second moment $\mu_2$ of fluctuation in $\Delta J_\mu$
spectra for the near-yrast $n$p-$n$h bands. The lowest twenty bands
are sampled from each nucleus. The average
value of the $\mu_2$'s calculated for the ten nuclei (see text) is
plotted with closed circles.
The bars represent the standard deviation of $\mu_2$'s.
The random limit for the average $\mu_2$  is
indicated by the solid line and
the thick bar indicate the limit standard deviation.

\item{Fig.4} The density of levels calculated in the extended
cranked shell model for the levels with $(-,1)$ in ${}^{168}$Yb
at $\omega=0.5$ MeV. Results with three different truncations
(number of basis states 600, 1000, and 1400) are plotted.
The Fermi-gas formula \nref{11} with the  level density parameter
$a=15.3$ MeV${}^{-1}$ is also plotted with a dashed curve
for comparison.

\item{Fig.5} The calculated rotational E2 strength distribution
as a function of the $\gamma$-ray energy $E_\gamma$. The four
figures correspond to the transitions decaying from the first,
6-th, 24-th, and 200-th levels with $(-,1)$ in
${}^{168}$Yb at $\omega=0.5$ MeV, respectively.
The magnitude of the strength is given as
the fraction relative  to the
total strength.

\item{Fig.6} The rotational strength distributions are shown for
transitions from 50 levels averaged in each bin.
The left-top quarter is for the first bin (from the first
to the 50-th level) with $(-,1)$.
The other three quarters
are for the levels from 51-th to 100-th, 101-th to 150-th and
151-th to 200-th.
The excitation energy from the yrast line is indicated for each bin.
The smoothed distribution function
is drawn with a solid curve.
The transitions shown are for $(-,1)$ in ${}^{168}$Yb
at $\omega=0.5$ MeV.

\item{Fig.7}
The distribution of the normalized rotational strengths $s_{ij}$
for ${}^{168}$Yb at $\omega =0.5$ MeV. For convenience
we draw the distribution as a histogram
against the square-root $\sqrt{s_{ij}}$ of the normalized strength.
In this representation the Porter-Thomas distribution, shown as
dashed curves,  exhibits Gaussian shape.

\item{Fig.8}
The distribution of the normalized rotational strengths.
In this figure, the transitions employed in the right-bottom quarter
of Fig.7 is divided into two groups according to whether their
transition $\gamma$-energies are in the central part of the
rotational strength function or in the tail part.

\item{Fig.9}
The branching number is shown as a function of the excitation
energy from the yrast. The histogram is the result of the
microscopic calculation while the dashed curve represents
the Porter-Thomas limit given by Eq.(12) in which
the rotational damping width $\Gamma_{rot}$ of 200 keV is adopted.
The values are calculated for ${}^{168}$Yb at $\omega=0.5$ MeV.

\item{Fig.10}
The $\Delta _3$ statistics of the energy level fluctuation
associated with the routhian eigenvalues
${e'_\alpha (\omega )}$ in
${}^{168}$Yb at $\omega =0.5$ MeV. The eigenvalues are
grouped into four bins in the same way as in Figs.6 and 7.
The long-dashed and the dashed curves represent
the limits of the Poisson distribution and the GOE random matrix model,
respectively.
The fitted level density for $(-,1)$
is shown in the inset with the adjusted parameters of the
constant temperature formula.

\item{Fig.11}
The E2 strength distributions as a function of the $\gamma$-ray
energy for the lowest 9 levels with $(\pi,\alpha)=(+,1)$ in
${}^{168}$Yb at $\omega=0.5$ MeV. The branching number
as well as the energy above yrast are also shown for each level.

\item{Fig.12}
The distribution of the normalized rotational strengths $s_{ij}$,
where the lowest 50 levels for each $(\pi,\alpha)$
are grouped into smaller bins of 10 levels. The approximate
energy regions corresponding to the bins are also shown.
In this figure the transitions in the 10 nuclei around ${}^{168}$Yb
are sampled.

\item{Fig.13}
Distribution of the normalized nearest neighbor level spacings $S$
for the unmixed rotational-band levels, defined by $n_{\rm branch}<2$,
near the yrast at $\omega=0.5$ MeV (the upper half).
The result for the independent-particle
cranked shell model without the residual interaction at the same $\omega$
is shown in the lower half.
The spacings are sampled from the ten nuclei around
${}^{168}$Yb and are normalized to give $<S>=1$.

\bye